\documentclass{aa}
\usepackage{graphicx}

\begin{document}
\title{The distance to the giant elliptical galaxy NGC~5128
\thanks{Based 
on observations collected at the European
Southern Observatory, Paranal, Chile, within the Observing Programmes
63.N-0229, 65.N-0164, 67.N-503, 68.B-0129 and 69.B-0292 and at 
La Silla Observatory, Chile, within the Observing Programme 64.N-0176(B).}}

\author{M. Rejkuba}

\offprints{M. Rejkuba}

\institute{European Southern Observatory, Karl-Schwarzschild-Strasse
           2, D-85748 Garching, Germany\\
           E-mail: mrejkuba@eso.org}

\date{1 July 2003 / 18 September 2003}
\titlerunning{Distance to NGC 5128}

\abstract{The distance to NGC~5128, the central galaxy of
the Centaurus group and the nearest giant elliptical to us, 
has been determined
using two independent distance indicators: the
Mira period-luminosity (PL) relation and the luminosity of the tip 
of the red giant branch (RGB). The data were taken at two 
different locations in the halo of 
NGC~5128 with  the ISAAC near-IR array  on ESO VLT. 
From more than 20 hours of observations with ISAAC a very deep 
$K_s$-band luminosity function was constructed. The tip 
of the RGB is detected at $\mathrm{K_s}=21.24 \pm 0.05$~mag. 
Using an empirical calibration
of the $K$-band RGB tip magnitude, and
assuming a mean metallicity of $[\mathrm{M}/ \mathrm{H}]=-0.4$~dex and
reddening of $\mathrm{E}(B-V)=0.11$, a distance modulus of NGC~5128 
of $(\mathrm{m}-\mathrm{M})_0=27.87 \pm 0.16$ was derived. The comparison of
the $H$-band RGB tip magnitude in NGC~5128 and the Galactic Bulge 
implies a distance modulus of NGC~5128 of 
$(\mathrm{m}-\mathrm{M})_0=27.9 \pm 0.2$ in 
good agreement with the $K$-band RGB tip
measurement.
The inner halo field has larger photometric errors, brighter
completeness limits and a larger number of blends. 
Thus the RGB tip feature is not as sharp as in the outer halo field. 
The population of stars above the tip of the RGB amounts to 2176 stars
in the outer halo field (Field~1) and 6072 stars in the inner halo
field (Field~2). The large majority of these sources belong to the
asymptotic giant branch (AGB)
population in NGC~5128 with numerous long period variables.
Mira variables were used to
determine the distance of NGC~5128 from a period-luminosity relation
calibrated using the Hipparcos parallaxes and LMC Mira period-luminosity
relation in the $K$-band. This is the first Mira period-luminosity relation 
outside the Local Group. A distance modulus
of $27.96 \pm 0.11$ was derived, 
adopting the LMC distance modulus of $18.50 \pm 0.04$.
The mean of the two methods yields a
distance modulus to NGC~5128 of 
$27.92 \pm 0.19$ corresponding to $D=3.84 \pm 0.35$ Mpc.
\keywords{Galaxies: elliptical and lenticular, cD --
          Galaxies: stellar content --
          Stars: fundamental parameters --
          Galaxies: individual: NGC~5128}
}

\maketitle

%
%
\section{Introduction}

Thanks to its proximity NGC 5128 (Centaurus A) has attracted lots of 
attention. 
It is the closest representative of the active radio galaxy and 
giant elliptical galaxy class of objects and is among the nearest 
AGNs. The latest review (Israel~\cite{israel98})
summarises its characteristics and it is a good starting point for
the vast literature about this galaxy. It also gives a summary of distance 
determinations, which started converging in the last decade from the early 
highly uncertain values ranging between 2.1 and 8.5 Mpc 
(Sersic~\cite{sersic58}, Sandage \& Tammann~\cite{sandage&tammann74}) towards
a less uncertain range between 3.2~Mpc (Hui et al.~\cite{hui+93}) and 4.2~Mpc 
(Tonry et al.~\cite{tonry+01}). 

The recent determinations of distance to NGC 5128 applied a range of methods.
Hui et al.~(\cite{hui+93}) used  the 
planetary nebula luminosity function to derive
a distance modulus of $(\mathrm{m}-\mathrm{M})_0=27.73 \pm 0.14$ mag. 
 The globular
cluster luminosity function was analysed by Harris et al.~(\cite{harris+88}) 
yielding $(\mathrm{m}-\mathrm{M})_0=27.53 \pm 0.5$ mag.
More recent globular cluster searches in this galaxy 
(e.g.~Rejkuba~\cite{rejkuba01}, Peng~\cite{pengPhD}) 
will allow a more precise determination of distance with 
this method through  a
much better sampled globular cluster luminosity function.

The luminosity of the red giant branch (RGB) 
tip stars in the $I$-band is a recognised distance 
indicator (e.g. Lee et al.~\cite{lee+93}). In NGC~5128 
it was first used by Soria et al.\ (\cite{soria+96}) who resolved the stellar
halo using HST+WFPC2. They derived  a distance modulus of 
$(\mathrm{m}-\mathrm{M})_0=27.86 \pm 0.16$ mag for WF chips and 
$(\mathrm{m}-\mathrm{M})_0=27.76 \pm 0.16$ mag for  the 
PC chip, and adopted a mean
distance modulus of $(\mathrm{m}-\mathrm{M})_0=27.8 \pm 0.2$ mag. More 
recently, Harris et al.~(\cite{harris+99}) used deeper HST+WFPC2 photometry
in a less crowded field to derive  the 
distance. Their RGB tip luminosity analysis 
resulted in $(\mathrm{m}-\mathrm{M})_0=27.98 \pm 0.15$ mag. Moreover, the same 
authors adjusted  the Hui et al.\ (\cite{hui+93}) distance modulus to 
$(\mathrm{m}-\mathrm{M})_0=27.97 \pm 0.14$, increasing it by 0.2 mag 
in order to correct for the contemporary Local Group distance scale and 
the M31 distance of $(\mathrm{m}-\mathrm{M})_{M31}=24.5$
(van den Bergh \cite{vandenbergh95}, Harris \cite{harris99}).

More than  the planetary nebula luminosity function,  the 
surface brightness 
fluctuations (SBF) method received several revisions of its calibration 
zero point. Tonry \& 
Schechter~(\cite{tonry&schechter90}) first derived 
$(\mathrm{m}-\mathrm{M})_0=27.48 \pm 0.06$ using I-band SBF.
This value was subsequently revised to 
$(\mathrm{m}-\mathrm{M})_0=27.71 \pm 0.10$ by Israel 
(\cite{israel98}) who used 
the results from Tonry (\cite{tonry91}) and then to 
$(\mathrm{m}-\mathrm{M})_0=28.18 \pm 0.07$ by Marleau et al.\ 
(\cite{marleau+00}) after Tonry et al.\ (\cite{tonry+97}). Most recently, 
Tonry et al.\ (\cite{tonry+01}) report $I$-band SBF
yielding a distance modulus of $(\mathrm{m}-\mathrm{M})_0=28.12 \pm 0.15$. In 
their list of nearby galaxies with SBF distance measurements, 
NGC~5128 occupies $9^{th}$ place and is the nearest giant elliptical galaxy.
The distance modulus determinations from 
the literature are summarized in Table~\ref{tabdist1}.

\begin{table*}
\centering
\caption[]{Summary of the distance distance modulus (DM) determinations 
from literature.}
\label{tabdist1}
\begin{tabular}{rcll}
\hline
\hline
\# &DM (mag) & Method & Reference \\
\hline
\hline
1&26.6      & Stellar luminosity function (LF) & Sersic~\cite{sersic58}\\
2&29.6      & Largest HII regions & Sandage \& Tammann~\cite{sandage&tammann74}\\
3&$27.73 \pm 0.14$& Planetary nebula LF & Hui et al.~\cite{hui+93}\\
4&$27.97 \pm 0.14$& Planetary nebula LF & (3) revised by Harris et al.~\cite{harris+99}\\
5&$27.53 \pm 0.5$ & Globular cluster LF & Harris et al.~\cite{harris+88}\\
6&$27.86 \pm 0.16$& $I$-band RGB tip (WF chips of WFPC2)& Soria et al.~\cite{soria+96}\\
7&$27.76 \pm 0.16$& $I$-band RGB tip (PC chip of WFPC2) & Soria et al.~\cite{soria+96}\\
8&$27.98 \pm 0.15$& $I$-band RGB tip (WFPC2) & Harris et al.~\cite{harris+99}\\
9&$27.48 \pm 0.06$& $I$-band SBF & Tonry \& Schechter \cite{tonry&schechter90}\\
10&$27.71 \pm 0.10$& $I$-band SBF & (9) revised by Israel \cite{israel98} \\
11&$28.18 \pm 0.07$& $I$-band SBF & (9) revised by Marleau et al.~\cite{marleau+00}\\
12&$28.12 \pm 0.15$& $I$-band SBF & Tonry et al.~\cite{tonry+01}\\
\hline
\end{tabular}
\end{table*}

In this paper I use Mira variables from the long period variable star 
catalogue in NGC~5128 (Rejkuba et al.~\cite{rejkuba+03}) 
and the $H$ and $K$-band luminosity 
functions to derive independent measurements of distance to NGC~5128. 
Data are briefly described in Sect.~\ref{data}, and in Sect.~\ref{LF} 
$J$, $H$, and $K$-band luminosity functions are analysed.  The 
distance to
NGC~5128 is derived from the RGB tip magnitude in Sect.~\ref{RGBtip}.
 The 
Mira period-luminosity (PL) diagram is constructed in Sect.~\ref{MiraPL} and 
used to determine  the distance to NGC~5128. 
Finally the results are 
summarized in Sect.~\ref{conclusions}.

%
%
\section{The data and photometry}
\label{data}

Data used here were described in detail in Rejkuba et al.\ 
(\cite{rejkuba+01} and \cite{rejkuba+03}). They consist of a set of 20 
$K_s$-band epochs, 1 $J_s$, and 1 $H$-band image of  a field 
located  $\sim 17\arcmin$ 
north-east from the center of NGC 5128 (Field~1). The second field (Field~2) 
was observed once in  the 
$J_s$ and $H$-bands and 24 times in  the $K_s$-band. 
It is located 
$\sim 9\arcmin$ south from the galactic center. All the observations, except
one $K_s$-band epoch of Field~2, 
were taken with  the 
ISAAC near-IR array at UT1 (Antu) telescope at ESO Paranal
Observatory in service mode. The additional $K_s$-band observation of Field~2 
was secured during an observing run with  the SOFI near-IR array 
 on the NTT 
at ESO La Silla Observatory under exceptional seeing conditions.

For the details about the photometry and completeness simulations the reader
is referred to the above mentioned papers. The photometry is complete more than
50\% in the $K_s$ and $H$-bands 
for stars brighter than 22.5 for Field~1 and 21.6 for Field~2. 
The 50\% completeness 
limit for  the 
$J_s$-band is 23.25 and 22.5 mag for Field~1 and 2, respectively.

Multi-epoch $K_s$-band observations span a 1197 day interval and were used to
search for variable stars. A total of 1504 red variables were detected in 
the two halo fields. For 1146 variables with at least 10 good measurements, 
periods, amplitudes and mean $K_s$ magnitudes were determined using
Fourier analysis and non-linear sine-curve fitting algorithms. Almost all these
variables belong to the class of long period variables (LPVs) with Mira and
semiregular variable stars. The complete catalogue of all LPVs is presented in
Rejkuba et al.\ (\cite{rejkuba+03}) to which reader is referred for discussion
of completeness and accuracy of period and amplitude determination. 

\begin{figure}
\centering
\includegraphics[width=7cm,angle=270]{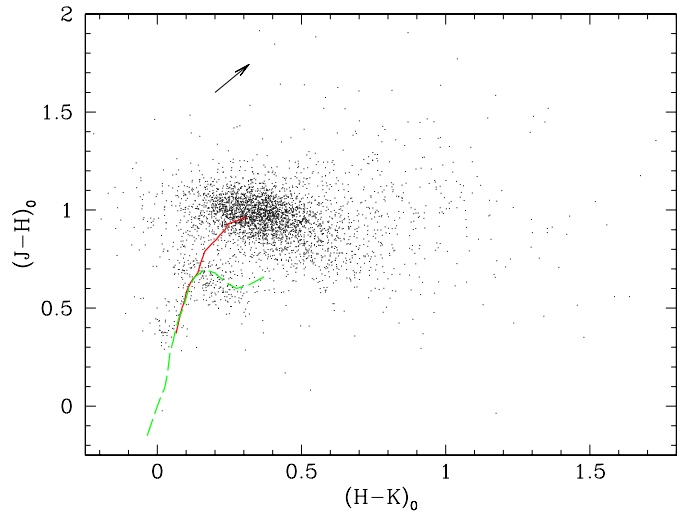}
\caption[]
	{Color-color diagram for all the stars in Field~1 and 2 with
	ALLFRAME photometry errors smaller than 0.1 mag. Foreground
	Galactic dwarf and giant stars follow the long-dashed and solid 
	lines indicating their intrinsic colors (Bessell \& Brett
	\cite{bessell+brett88}). A reddening vector corresponding 
	to $\mathrm{E}(B-V)=0.5$ mag is shown.
	}
\label{color-color}
\end{figure}

The ISAAC filters used are $J_s$, $H$ and $K_s$. The $J_s$ filter was 
preferred over $J$, due to the red leakage of the latter. It is centered at 
1.24 $\mu$m and has a width of 0.16 $\mu$m. The transformation to the $J$-band
of the LCO photometric system 
(Persson et al.~\cite{persson+98})  was obtained through the 
following transformation (Chris Lidman private communication):
\begin{equation}
J_{LCO} = Js_{ISAAC} + 0.033*(Js-Ks)_{ISAAC} - 0.022
\label{J_ISAAC_LCO}
\end{equation}
The ISAAC $H$ and $K_s$ filters are on the LCO photometric system. 

It should be noted
that to all the $K_s$-band magnitudes from the 
catalogue (Rejkuba et al.~\cite{rejkuba+03}) a constant of 0.1 mag has 
been subtracted. 
A comparison of the color-color diagrams with the intrinsic colors of Galactic
dwarfs and giants (Bessell \& Brett~\cite{bessell+brett88}) has revealed
an error in the aperture correction of the $K_s$-band photometry. For  
reference the $J-H$ vs. $H-K$ color-color diagram is shown in 
Fig.~\ref{color-color} with the 
Bessell \& Brett (\cite{bessell+brett88}) fiducials overplotted. 
The solid line indicates the 
intrinsic colors of early to late type giants and  the 
long-dashed line
is for dwarfs. Foreground Galactic stars are found along these lines. The 
large majority of the stars in NGC~5128 are found close to the locus of late
type giants and redwards from there, where long period variables and carbon
stars are located.
NGC~5128 photometry has been de-reddened adopting $\mathrm{E}(B-V)=0.11$
(Schlegel et al.~\cite{schlegel+98}) and 
the Cardelli et al.~(\cite{cardelli+89}) reddening
law and it has been transformed to the Bessell \& Brett 
(\cite{bessell+brett88}) $JHK$ photometric system using the transformation
equations available from the 2MASS web page\footnote{\tt 
http://www.astro.caltech.edu/$\sim$jmc/2mass/v3/transformations/}.

%
%
\section{Luminosity functions}
\label{LF}

\begin{figure}
\centering
\includegraphics[width=7cm,angle=270]{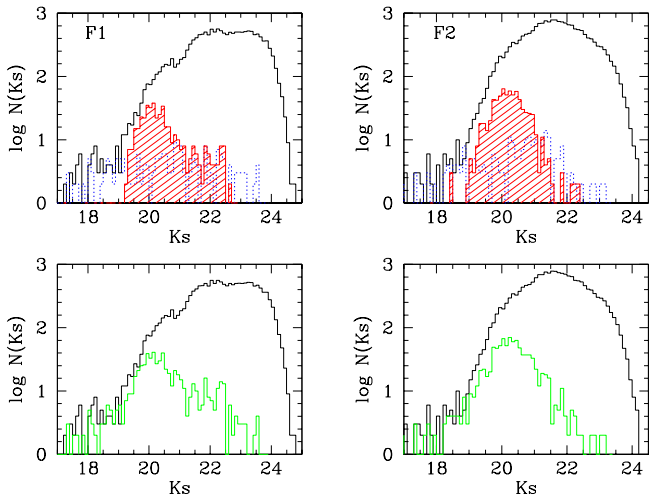}
\caption[]
	{$K_s$-band luminosity function for all stars (solid black line) 
	in Field~1 (left) and Field~2 (right).
	Top:  the dotted line histogram is for stars 
	bluer than $-(V-K_s)+25$ (see Rejkuba et al.~\cite{rejkuba+01}). 
	Mostly they are foreground Galactic stars as well as young
	blue and red supergiants in Field~1. The
	variable star luminosity function (red shaded histogram)
	is also compared with the complete luminosity function.
	Bottom: A sum of the foreground and variable star
	luminosity function is compared to the total luminosity function
	of the two fields.
	}
\label{lfkall}
\end{figure}

The $K_s$-band luminosity functions for all the stars in Field~1 (left) 
and 2 (right) are shown in
Fig.~\ref{lfkall} as solid line histograms. These are not corrected for
incompleteness nor for Galactic foreground contamination. The number of
foreground Milky Way stars can be obtained from the color-color 
diagram (Fig.~\ref{color-color}) or from optical-near-IR 
color-magnitude diagrams published by Rejkuba et al.\ (\cite{rejkuba+01}).
There are 194 stars with near-IR colors consistent with dwarf or early type
giants in  the Field~1 color-color diagram. Similarly, $192$ sources have 
blue colors in  the $VK_s$ CMD
(see Rejkuba et al.~\cite{rejkuba+01}). They are plotted as a 
dotted histogram in Fig.~\ref{lfkall}. These stars are mostly
foreground Galactic stars as well as young blue and red supergiants in
Field~1.  The variable star luminosity function is plotted as a shaded 
histogram. It 
is clear that majority of the variables have luminosities brighter
than the RGB tip (see below) 
and are in the thermally pulsing asymptotic giant branch (TP-AGB) 
phase. In the lower panels
the total luminosity functions are compared with the luminosity functions of
the sums of the contributions of foreground stars and LPVs.

There are $2176$ stars in Field~1 above the tip of the RGB 
($K_s<21.24$, see next section) detected in at least 3 
$K_s$-band images. $198$ of these sources have not been detected 
in $J_s$ and/or $H$-band frames.
According to the $VK_s$ CMD, 192 stars
have blue colors and they belong either to the young population
in Field~1 or to the Galactic foreground. Among the remaining
1984 stars 426 are variable stars 
(accounting for 71\% of variables detected 
in this field).

In Field~2 there are 6072 sources brighter than $K_s=21.24$ detected in at 
least $3$ $K$-band images.
$351$ of these sources have not been detected in $J$ and/or $H$-band frames.
Of these $217$ objects have blue colors 
and belong to the Galactic foreground contamination. 
Of the remaining $5855$ stars brighter than the RGB tip $829$ are
asymptotic giant branch (AGB) 
variables (accounting for 92\% of detected variables in this field). 

\begin{figure*}
\centering
\includegraphics[width=12.0cm,angle=270]{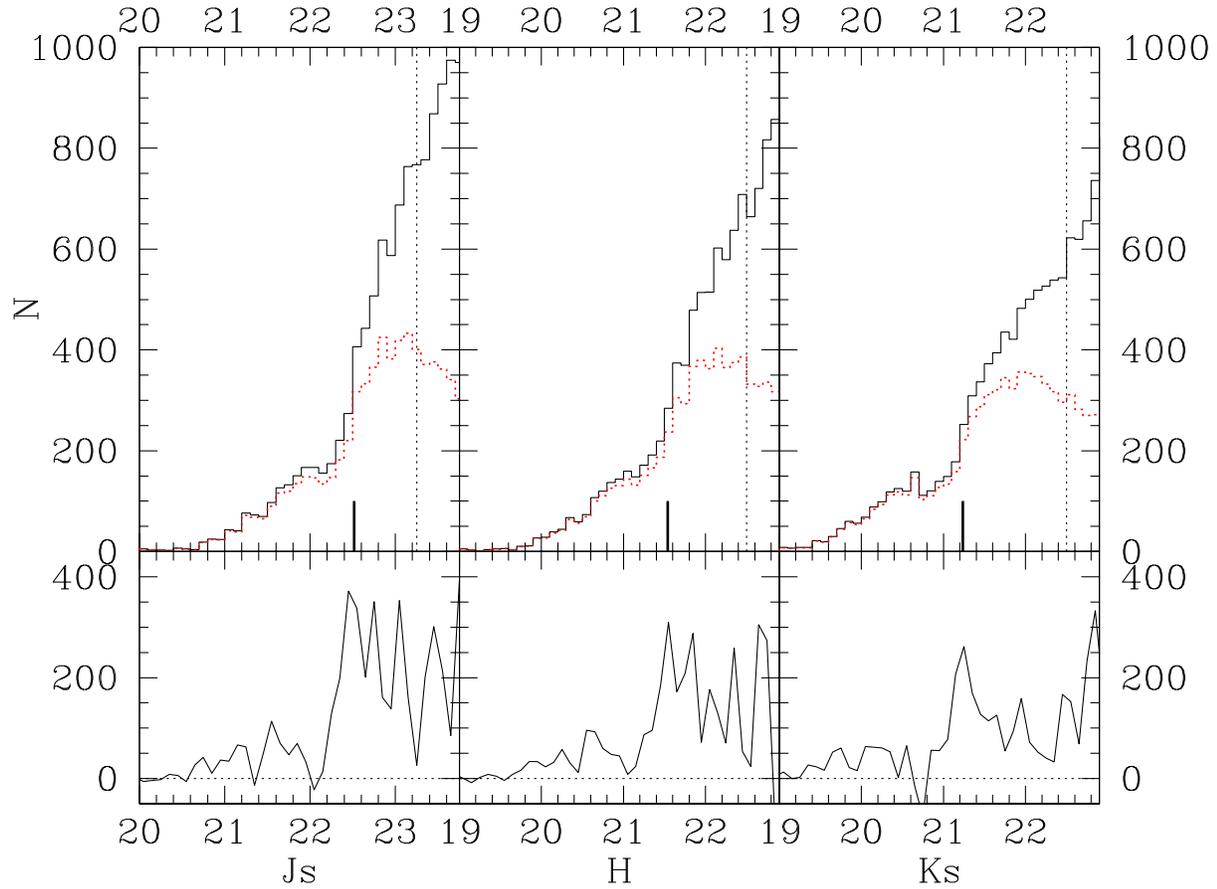}
  \caption[]
	{$J_s$, $H$ and $K_s$-band luminosity functions for Field~1 
	are shown in the top panels. Dotted lines are used for the 
	observed and solid lines for 
	the completeness-corrected luminosity functions.
	In the bottom panels the Sobel edge-detection filter
	response curves are displayed. Vertical dotted lines indicate 50\%
	completeness limits and the short thick marks indicate RGB tip
	magnitudes measured in this Field.}
  \label{lfk1}
\end{figure*}  
\begin{figure*}
\centering
\includegraphics[width=12.0cm,angle=270]{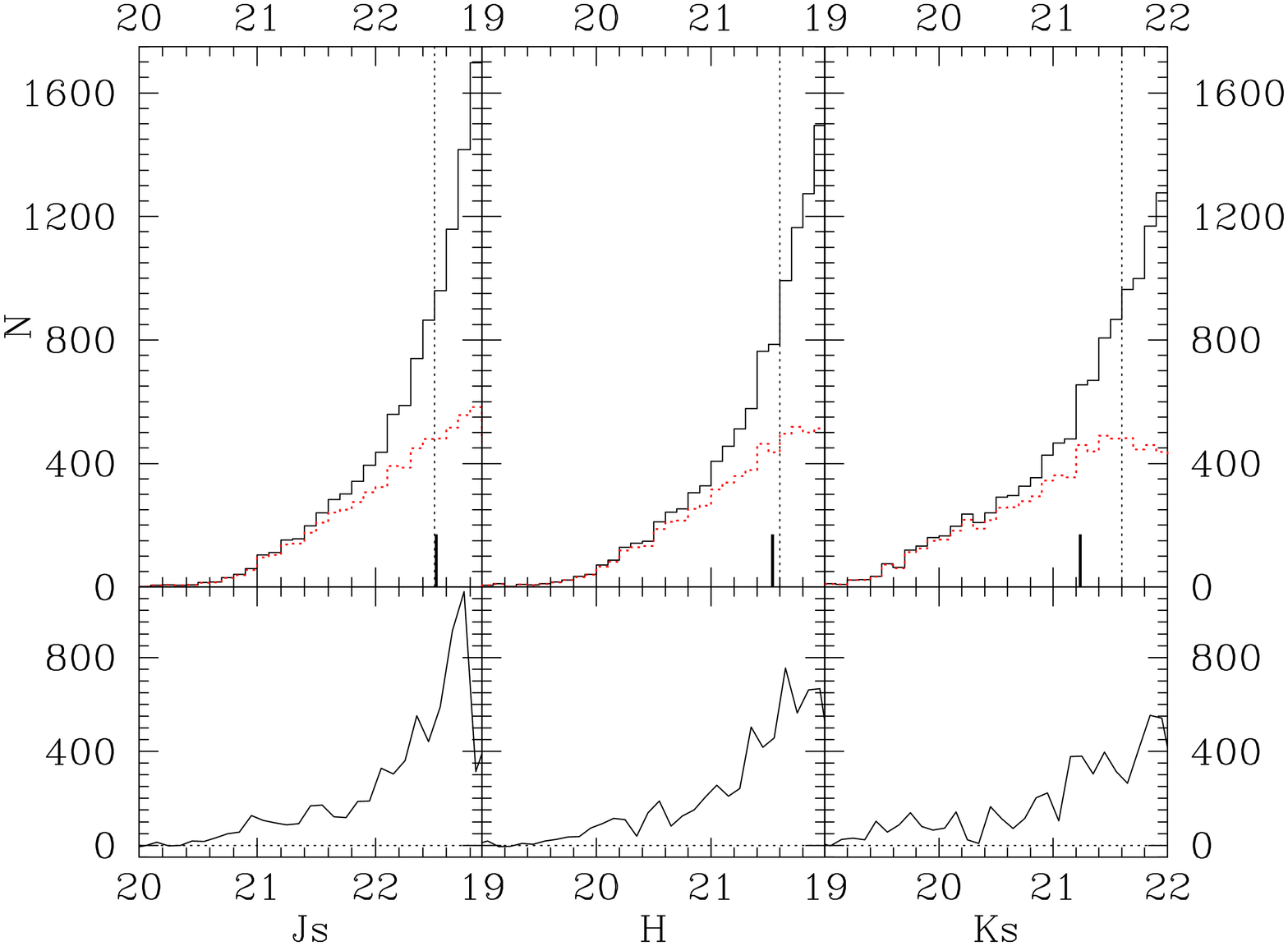}
  \caption[]
	{$J_s$, $H$ and $K_s$-band luminosity functions for Field~2 
	are shown in the top panels. Dotted lines are used for the 
	observed and solid lines for 
	the completeness-corrected luminosity functions.
	In the bottom panels the Sobel edge-detection filter
	response curves are displayed. Vertical dotted lines indicate 50\%
	completeness limits. Short thick marks indicate RGB tip magnitudes
	measured in Field~1 (Fig.~\ref{lfk1}).}
  \label{lfk2}
\end{figure*}  

Some of the bright non-variable AGB giants could be blends of 
2 RGB tip stars. Renzini (\cite{renzini98}) provided the equations
that enable the calculation of the expected number of
stars in a given evolutionary phase ($N_j$) 
in an image in which the total sampled luminosity is $L_T$:
\begin{equation}
N_j = B(t) L_T t_j
\end{equation}
$B(t)$ is the specific evolutionary flux of the population or the number
of stars entering or leaving any post main sequence evolutionary stage per
year and per solar luminosity of the population. It is a very weak
function of age and ranges from $\sim 0.5\times10^{-11}$
to $\sim2.2\times10^{-11}$ stars~L$_\odot^{-1}$~yr$^{-1}$,
as age increases from 10~Myr to 15~Gyr. $t_j$ is the duration of the
evolutionary phase. 
For a solar-metallicity, 15~Gyr old population
$L_T \simeq 0.36 L_K$.
We can estimate the total sampled luminosity
in the two fields by summing the observed counts on the
dark subtracted, flat-fielded, sky-subtracted and calibrated frames. 
Using the combined frames for the reference epoch observed on JD=2451734, 
the total sampled luminosity 
is $5\times 10^7$~L\sun~ and $7\times 10^7$~L\sun~ for Field~1 and
2, respectively.
From the surface brightness profile of Dufour et
al.~(\cite{dufour+79}) the total luminosities are $5.9$ and
$28.7\times10^7$~L\sun, while from that of Mathieu et
al.~(\cite{mathieu+96}) they are $9.5$ and $31 \times 10^7$~L\sun. Our
measured value for Field~1 is close to that of Dufour et al., but
the measured luminosity of Field~2 is much lower. 

From Table~1 of Renzini (\cite{renzini98}) the expected number of LPVs in 
an old near-solar metallicity population is $\sim 5/10^6$~L\sun. The total
sampled luminosity of Field~1 is 50-100 times higher, yielding 250-500
LPVs in agreement with the 437 LPVs detected in that field 
(Rejkuba et al.~\cite{rejkuba+03}). 
In Field~2 our measured luminosity would suggest $\sim350$ LPVs, 
but the surface brightness measurements from the literature predict 
$\sim 1500$ LPVs. For comparison, we have detected 903 variable stars and
measured periods for 709 LPVs in Field~2.

The probability that a pixel contains two stars in phase $j$ is $\sim
N_j^2$ and the number of 2-star blends in a frame with a total
number of pixels $N^{pix}$ is:
\begin{equation}
N_{2j}=N_j^2N^{pix}=[B(t) L_T t_j]^2/N^{pix}
\end{equation}
The resolution of ground based images is worse
than 1 pixel and the sampled luminosity has to be divided
by the available number of resolution elements. Each
resolution element is $0\farcs31$ for Field~1 and 
$0\farcs36$ for Field~2, corresponding to the best seeing epochs used to
detect sources. 
Hence the expected number of blends of two RGB tip giants in Field~1 is
between 100 and 400 for the total sampled luminosity between 5 and $9.5 
\times 10^7$~L\sun. In Field~2 it is $\sim 250$ if the measured luminosity
from the image is assumed, but it rises to $4900$ for 
$31 \times 10^7$~L\sun~
from Mathieu et al. The expected number of blended stars in the 
vicinity of the RGB tip is low enough to measure the RGB tip discontinuity 
in the luminosity function of Field~1, but may smear out that feature 
in Field~2.
Taking the highest probable number of blends, there are at least 1150
non-variable bright AGB giants in Field~1 and some 150 in Field~2. Rejkuba
et al.\ (\cite{rejkuba+03}) made extensive simulations and 
estimated the completeness of the detection of LPVs as a function of their
magnitude, period and amplitude. These simulations show a similar 
completeness level of the LPV catalogue in the two fields 
implying that the number
of bright extended giant branch non-variable stars is larger in Field~1.
However, these stars could be semi-regular small-amplitude ($\Delta K \la 0.3$)
variables for which the LPV catalogue is less complete. 
Their near-IR properties
will be discussed together with near-IR properties of the LPVs in
a forthcoming paper (Rejkuba et al.~in  preparation).

LPVs account for at least 26\% and 70\% 
of the extended giant branch population in
Fields 1 and 2, respectively. Their high luminosity 
cut-off extends beyond $\mathrm{M}_\mathrm{K}=-8.7$. 
The maximum brightness achieved by
AGB stars is evidence for the presence of an 
intermediate-age population. 

%
%
\section{NGC~5128 distance}
\label{distance}

In this section  the distance to
NGC~5128 is measured using two different methods: 
(i) the brightness of the RGB tip in the
$K$ and $H$-bands and (ii) the Mira period-luminosity relation.

\subsection{RGB tip}
\label{RGBtip}

The $J_s$, $H$ and $K_s$-band luminosity functions of Field~1 stars 
are presented in
Fig.~\ref{lfk1}. These are the deepest luminosity functions for a
stellar population belonging to a galaxy beyond the Local Group. 
The solid lines are used for completeness-corrected luminosity functions 
and dotted lines for the observed ones. Vertical dotted line indicates 50\%
completeness limit of the photometry. Short thick marks indicate the
measured RGB tip magnitudes in Field~1.

In the bottom panels of Fig.~\ref{lfk1} 
I show the luminosity functions convolved with the 
Sobel edge detection filter with a $(-2,0,+2)$ kernel 
(Lee et al.~\cite{lee+93}). 
They are used as localised
slope estimators with 2-point smoothing.

There is a very sharp peak
in the Field~1 response function curve at $J=22.54 \pm 0.06$, 
$H =21.54 \pm 0.06$, and $K_s=21.24\pm0.05$.
For easier comparison with the models and literature, the
$J$-band measurement is transformed to the LCO photometric system according to 
Eqn.~\ref{J_ISAAC_LCO}. 
These maxima appear at the position where there is a strong change in the 
slope of the RGB luminosity function and it corresponds to the 
magnitude of the tip of the RGB. The error in the RGB tip magnitude is 
obtained by averaging over 100 different measurements of Sobel kernel 
peak magnitude, where luminosity 
function sampling is varied by 0.001 mag over 0.08 to 0.2 mag.

The RGB tip feature is smoothed over a couple 
of tenths of magnitude in Field~2 (Fig.~\ref{lfk2}). 
From the 2-point Sobel kernel,  the RGB tip in Field~2 is centered 
around $K_s=21.22 \pm 0.15$ mag. 
The broadening and smoothing of the peak
is due to much shallower completeness limits (note the short thick marks
indicating the RGB tip magnitude measured in Field~1), resulting in 
larger photometric errors, as well as due to contribution
of blends, the presence of AGB stars and  a larger metallicity spread in
this field. 

The tip of the RGB in the $K$-band is a function of metallicity. Its
dependence has been measured empirically by Ferraro et
al.~(\cite{ferraro+00}) using a set of Galactic globular clusters with a
range of metallicities. They obtained the following relation:
\begin{equation}
{\rm M}_{\rm K}^{\rm TRGB} = -(0.64 \pm 0.12){\rm [M/H]} - (6.93 \pm 0.14)
\end{equation}

The mean metallicity for the stars in  the 
NGC~5128 halo has been inferred 
to be around $\mathrm{[M/H]}=-0.4$~dex 
(Marleau et al.~\cite{marleau+00}, Harris \& Harris~\cite{harris+harris00}, 
Rejkuba \cite{rejkubaPhD}). 
With this estimate, and assuming a reddening of
$\mathrm{E}(B-V)=0.11$ (Schlegel et al.~\cite{schlegel+98}), corresponding to
${\rm A}_{\rm K}=0.039$ (Cardelli et al.~\cite{cardelli+89}), 
the empirical calibration of the RGB tip $K$-band magnitude can be used 
to derive the distance to NGC~5128. The difference between the photometric 
bands is of the order of 0.01 mag (Persson et al.~\cite{persson+98}).
This yields a distance modulus of $(\mathrm{m}-\mathrm{M})_0=27.87 \pm 0.16$. 
The quoted error does not include the uncertainty in reddening and in mean
metallicity of the stars. A higher mean metallicity by 
0.1~dex implies a larger distance modulus by 0.06 mag.  The 
$K$-band is rather
insensitive to extinction, and reddening as high as 
$\mathrm{E}(B-V)=0.35$ would lower the distance modulus by 0.088 mag.

The above measurements can be compared to that of the RGB tip brightness 
in the LMC and in the Galactic Bulge. For example, 
Cioni et al.~(\cite{cioni+00}) have measured 
$K_s^{TRGB}(LMC)=11.94 \pm 0.04$ and $J^{TRGB}(LMC)=13.06\pm 0.02$; these
values are not corrected for metallicity. The 
difference between the RGB tip magnitudes in the LMC and NGC~5128 imply 
distance moduli differences of 
$\Delta(\mathrm{m}-\mathrm{M})_{K_{s}} = 9.26 \pm 0.06$~mag and
$\Delta(\mathrm{m}-\mathrm{M})_J = 9.39 \pm 0.06$~mag. Adopting a distance 
modulus of 18.5 for the LMC (Alves et al.~\cite{alves+02}) 
yields distance moduli for NGC~5128 of 
$27.76$ and $27.89$ using $K_s$ and $J$ bands, respectively. 
While this is within errors consistent with the above 
quoted $27.87 \pm 0.16$,  it should be remembered that 
the higher metallicity of the NGC~5128 stars with respect to the LMC, makes
the RGB tip magnitude brighter in $K$-band 
(e.g.~Zoccali et al.~\cite{zoccali+03}), hence the relative difference of 
distance modulus is a lower limit in  the $K_s$-band. 

The $J$-band RGB tip magnitude
is expected to decrease slowly as the metallicity increases, while the tip
luminosity in the $H$-band appears to be fairly constant in the 
metallicity range $-1.0\la \mathrm[Fe/H] \la 0.0$ 
(Zoccali et al.~\cite{zoccali+03}). Hence the latter could be used for 
distance determination. Schulte-Ladbeck et al.~(\cite{schulte-ladbeck+99}) 
have investigated the behaviour of  the 
$H$-band tip luminosity with age and
metallicity. They recommend $M_H^{TRGB} = -5.5\pm 0.1$ for 
$-2.3 <\mathrm{[Fe/H]} < -1.5$. Bertelli et al.~(\cite{bertelli+94}) models
suggest   $M_H^{TRGB} \sim -6$~mag for higher metallicity, while 
$M_H^{TRGB}= -6.4 \pm 0.2$ in the Galactic Bulge 
(Zoccali et al.~\cite{zoccali+03}). The latter yields a distance modulus to
NGC~5128 of $27.9 \pm 0.2$. Due to high extinction, small number of stars and possible 
differential reddening in the Galactic Bulge field observed, the
$M_H^{TRGB}$ 
value is rather uncertain. However, it yields  a
distance modulus to NGC~5128 of
$27.9 \pm 0.2$ in excellent agreement with that derived from the $K$-band 
RGB tip magnitude.

\begin{figure}
\centering
 \includegraphics[width=7.0cm,angle=270]{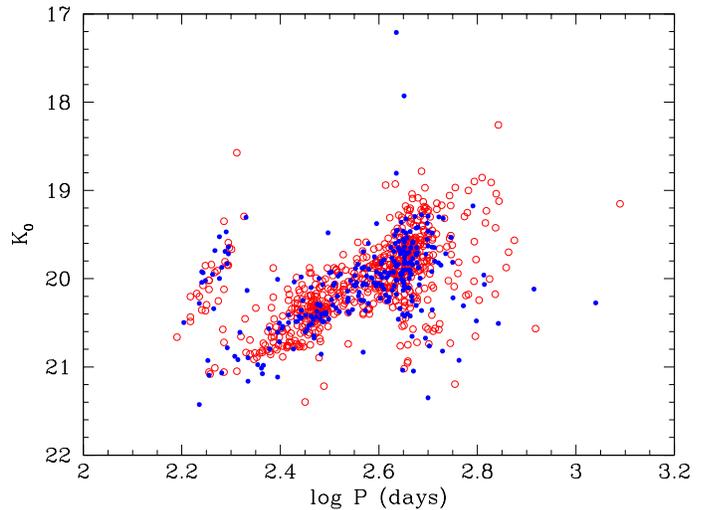}
   \caption[]
	{Period-luminosity diagram for all variables in NGC~5128 with 10
	or more data points with well determined periods from the LPV
	catalogue of Rejkuba et al.~(\cite{rejkuba+03}).
	Field~1 variables are plotted with filled and variables in Field~2
	with open symbols.}
  \label{plall}
\end{figure}  

\subsection{Mira PL relation}
\label{MiraPL}

In Fig.~\ref{plall} I plot the mean magnitudes $vs.$ logarithm of
periods for all the variables with 10 or more data points for which
derived periods had a significance parameter from Fourier analysis $<0.7$
(Rejkuba et al.~\cite{rejkuba+03}).
Field~1 variables are shown with filled and Field~2 with open symbols. 
The mean magnitudes are corrected for extinction and transformed to the
SAAO photometric system (Carter \cite{carter90}) in which most Mira studies
are published. 

\begin{figure}
\centering
\includegraphics[width=7cm,angle=270]{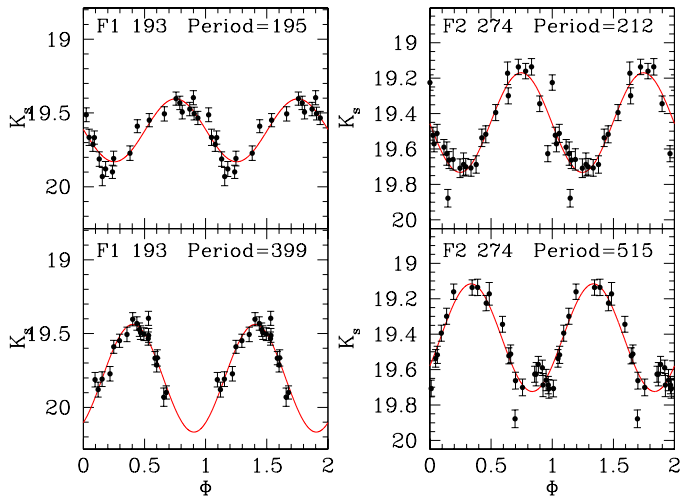}
  \caption[]
	{Example of two LPVs where aliasing is present. Both periods, a short
	one of $\sim 1/2$~yr and $\sim$~twice as long one 
	can be equally well fit.}
  \label{Palias}
\end{figure}  

\begin{figure}
\centering
 \includegraphics[width=7.0cm,angle=270]{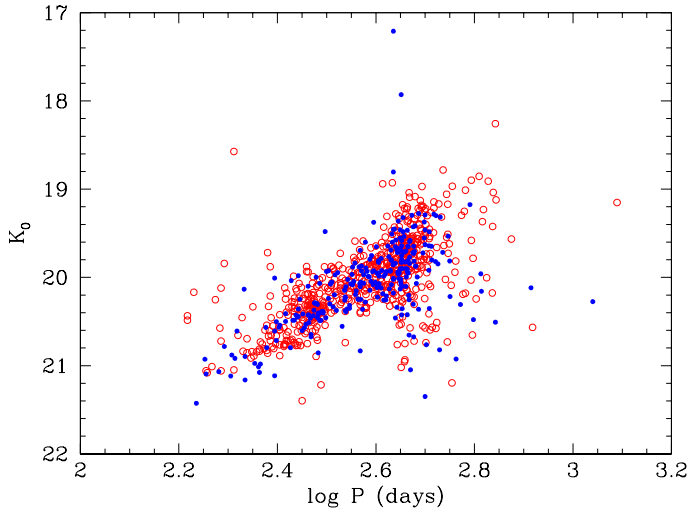}
   \caption[]
	{Period-luminosity diagram for all variables in NGC~5128 with 10
	or more data points with well determined periods. Alternative
	aliasing periods as listed in Table~\ref{Palias.tab} are plotted.
	Field~1 variables are plotted with filled and variables in Field~2
	with open symbols.}
  \label{plall_new}
\end{figure}  

\begin{table*}
\centering
\caption[]{LPVs with possible aliasing periods. Original field 
	identification and ID number, period, semi-amplitude, 
	reduced $\chi^2$ of the sine-curve fit, significance of the 
	original period and the mean $K_s$ magnitude are listed 
	in columns 1 to 6. The new, aliasing, period, semi-amplitude, reduced
	$\chi^2$ of the sine-curve fit and the associated
	average $K_s$ magnitude, are listed in columns 7, 8, 9 and 10,
	respectively.}
\begin{tabular}{crrrrrrrrr}
\hline
\hline
ID    & $P_{orig}$& $A_{orig}$&$\chi^2_{orig}$& $signif$& $<K_s>_{orig}$& $P_{new}$&
$A_{new}$&$\chi^2_{new}$&$<K_s>_{new}$ \\ 
\hline
\hline
F1 166  &214.  &  0.254&    4.30 &  0.340&  19.46&	522.&    0.262&    4.49&    19.44\\
F1 193	&195.  &  0.213&    2.30 &  0.140&  19.62&	399.&    0.365&    2.05&    19.80\\
F1 201	&189.  &  0.296&   10.90 &  0.390&  19.67&	382.&    0.667&    5.86&    20.07\\
F1 269	&197.  &  0.309&    4.20 &  0.230&  19.79&	401.&    0.549&    1.90&    20.10\\
F1 283	&185.  &  0.178&    3.80 &  0.660&  19.83&	372.&    0.377&    3.62&    20.07\\
F1 286  &197.  &  0.247&    9.20 &  0.550&  19.87&	403.&    0.444&    5.29&    20.10\\
F1 296	&194.  &  0.217&    2.50 &  0.260&  19.84&	392.&    0.413&    1.49&    20.08\\
F1 353	&652.  &  0.378&   56.00 &  0.630&  20.26&	652.&    0.861&   16.00&    20.35\\
F1 375	&196.  &  0.272&    4.70 &  0.300&  19.98&	400.&    0.498&    2.46&    20.26\\
F1 393 	&191.  &  0.220&    6.70 &  0.540&  20.02&	387.&    0.466&    6.61&    20.29\\
F1 425	&174.  &  0.242&    4.00 &  0.220&  20.07&	344.&    0.416&    2.65&    20.29\\
F1 431	&183.  &  0.226&    3.70 &  0.440&  20.10&	367.&    0.485&    2.24&    20.39\\
F1 441	&175.  &  0.243&    3.40 &  0.250&  20.08&	346.&    0.360&    1.77&    20.25\\
F1 522	&174.  &  0.237&    3.00 &  0.260&  20.19&	348.&    0.514&    2.76&    20.50\\
F1 527	&189.  &  0.180&    3.20 &  0.650&  20.15&	387.&    0.357&    1.62&    19.95\\
F1 536	&177.  &  0.287&    4.30 &  0.280&  20.17&	352.&    0.528&    1.96&    20.46\\
F1 793	&455.  &  0.077&   50.30 &  0.090&  20.56&	450.&    0.588&    6.60&    20.52\\
F1 816	&172.  &  0.220&    5.60 &  0.650&  20.44&	340.&    0.460&    4.39&    20.71\\
F1 908	&445.  &  0.299&    5.60 &  0.270&  20.55&	304.&    0.314&    5.41&    20.57\\
F1 1001	&184.  &  0.248&    2.60 &  0.310&  20.50&	370.&    0.422&    1.83&    20.27\\
F1 1089	&160.  &  0.238&    3.60 &  0.320&  20.65&	291.&    0.253&    3.68&    20.58\\
F1 1888	&445.  &  0.154&   16.70 &  0.170&  21.22&	202.&    0.325&   12.00&    21.30\\
F2 111  &486.  &  0.367&    3.20 &  0.140&  18.94&	545.&    0.367&    3.24&    18.94\\
F2 258	&193.  &  0.257&    4.20 &  0.450&  19.50&	413.&    0.330&    2.17&    19.58\\
F2 274	&212.  &  0.281&    4.60 &  0.090&  19.45&	515.&    0.304&    3.92&    19.42\\
F2 551  &730.  &  0.862&    6.70 &  0.030&  19.87&	240.&    0.450&    6.97&    19.89\\
F2 594	&197.  &  0.248&    3.80 &  0.080&  19.74&	402.&    0.440&    1.97&    19.98\\
F2 663  &198.  &  0.220&    1.80 &  0.040&  19.80&	400.&    0.390&    1.87&    20.02\\
F2 700	&200.  &  0.224&    2.80 &  0.130&  19.82&	419.&    0.325&    2.10&    19.97\\
F2 767  &722.  &  0.483&    5.10 &  0.270&  20.03&      363.&    0.432&    8.39&    19.96\\
F2 777  &194.  &  0.312&    8.30 &  0.240&  19.96&	399.&    0.546&    2.32&    20.23\\
F2 817  &177.  &  0.279&    5.50 &  0.210&  20.00&	352.&    0.575&    2.89&    20.32\\
F2 935	&197.  &  0.241&    4.80 &  0.220&  20.00&	401.&    0.451&    1.69&    20.25\\
F2 1044 &182.  &  0.242&    6.40 &  0.360&  19.99&	369.&    0.458&    5.61&    20.26\\
F2 1096	&185.  &  0.241&    4.50 &  0.190&  20.08&	371.&    0.538&    4.21&    20.42\\
%
F2 1216 &187.  &  0.306&    2.30 &  0.040&  20.06&	380.&    0.506&    1.28&    20.34\\
F2 1364 &180.  &  0.266&    9.10 &  0.620&  20.17&	364.&    0.359&    7.93&    20.31\\
F2 1378 &180.  &  0.235&    2.60 &  0.110&  20.17&	362.&    0.318&    3.04&    20.32\\
F2 1465 &175.  &  0.303&    6.60 &  0.180&  20.23&	343.&    0.334&    6.01&    20.30\\
F2 1588 &476.  &  0.129&   58.10 &  0.080&  20.41&	476.&    0.541&    5.34&    20.46\\
F2 1600 &180.  &  0.421&    7.00 &  0.220&  20.28&	367.&    0.510&    4.40&    20.10\\
F2 1623 &449.  &  0.305&   13.10 &  0.460&  20.42&	377.&    0.404&   11.61&    20.30\\
F2 1700 &460.  &  0.142&   34.50 &  0.080&  20.43& 	455.&    0.471&    7.32&    20.39\\
F2 1871 &165.  &  0.279&    4.00 &  0.260&  20.36&	328.&    0.398&    4.14&    20.18\\
F2 1917 &461.  &  0.192&   49.80 &  0.250&  20.49&	451.&    0.436&    6.39&    20.44\\
F2 2042 &172.  &  0.290&    3.50 &  0.110&  20.36&	340.&    0.434&    1.39&    20.17\\
F2 2097 &178.  &  0.495&   10.20 &  0.260&  20.45&	359.&    0.587&    7.38&    20.26\\
F2 2100 &450.  &  0.097&   36.00 &  0.300&  20.47&	433.&    0.486&    5.46&    20.34\\
F2 2287 &174.  &  0.299&    4.30 &  0.170&  20.51&	345.&    0.466&    3.73&    20.27\\
F2 2467 &177.  &  0.403&    6.40 &  0.230&  20.52&	351.&    0.455&    6.57&    20.42\\
F2 2554 &605.  &  0.281&   24.70 &  0.670&  20.52&	605.&    0.560&    9.27&    20.53\\
F2 3622 &155.  &  0.312&    3.80 &  0.160&  20.81&	267.&    0.328&    4.56&    20.86\\
\hline
\end{tabular}
\label{Palias.tab}
\end{table*}

All the variables with $K_0$ magnitudes brighter than $19.0$ have been 
carefully checked on the reference (best seeing) image. The two 
Field~1 LPVs with periods around 430 days,
both of which are brighter than $K_s<18.0$ 
are located in a highly crowded areas. 
The fainter of the two is actually blended with a background galaxy and the 
brighter might be a blend of more than two stars. Also the third 
brightest Field~1 LPV could be too bright due to
contributing light from the nearby stars. LPVs which are blended with 
nearby stars should have smaller amplitudes due to larger relative
contribution of the neighbour at minimum phases.
Only one of the eight 
Field~2 bright LPVs ($K_0<19.0$) is a possible blend of two stars. 

There are two sequences in the PL diagram in Fig.~\ref{plall}.
Multiple parallel sequences in the PL diagram of LPVs were 
reported by Wood \& Sebo (\cite{wood&sebo96}), Wood (\cite{wood00}) 
and Cioni et al.~(\cite{cioni+01})  
in the Magellanic Cloud. The more populated, longer period sequence 
corresponds
to sequence C of Wood (\cite{wood00}) where Mira variables are
found. The shorter period sequence B  discussed by Wood (\cite{wood00})
is approximately 1.3 mag brighter, while the shorter sequence in 
Fig.~\ref{plall} lies only 0.5-1 mag brighter. Moreover, it is is 
somewhat steeper, indicating a possible problem of aliasing periods. 

Actually, a close inspection, shows that most of the stars
on the shorter period sequence have periods close to
$1/2$ year and can be fitted similarly well with periods that are twice as 
long, and which place them on the Mira sequence. 
It is interesting to note that
this increases also the amplitude of the fit.
An example is shown in Fig.~\ref{Palias}. In the upper
panel the light curve is folded with the original, shorter, period and with 
the longer period on the bottom. 
Carefully phased additional observations of these variables would be necessary
to determine their periods unambiguously. The period distribution of
the LPVs and the comparison between the two fields is discussed by Rejkuba
et al.~(\cite{rejkuba+03b}).

In Table~\ref{Palias.tab} all the variables that could be fitted with
2 aliasing periods are listed. 
Columns~1 to 6 list: field identification and ID
number, original period, semi-amplitude, reduced $\chi^2$ of the sine-curve
fit, significance of the original period and the mean $K_s$ magnitude
obtained from the sine-curve fit. The new period, semi-amplitude, reduced
$\chi^2$ of the sine-curve fit with this new period and the associated 
average $K_s$ magnitude, are listed in columns 7, 8, 9 and 10,
respectively.
Additionally I have checked for possible aliasing periods all the LPVs that
are located below the Mira PL relation. In most cases no shorter period
could provide a satisfactory fit to the data, except for 2 
LPVs in Field~1 and 5 in Field~2. Fainter average $K$-band magnitudes
for the LPVs that lie below the Mira PL relation are due to extinction 
within their circumstellar shells (see also Kiss \& Bedding
\cite{kiss+bedding03}). 
In another 6 cases $\chi^2$ of the sine-curve fit was
improved and a new amplitude and mean magnitude  were 
calculated, although the
period remained very close to the original value. Apparently running the
sine-curve fitting algorithm in an automatic way failed to find the best
fitting amplitudes in a few cases. These are also 
listed in Table~\ref{Palias.tab}.
In the following distance determination these LPVs are not used. 
Fig.~\ref{plall_new} displays the PL diagram in which aliasing periods
($P_{new}$) from Table~\ref{Palias.tab} are used.

\begin{figure}
\centering
\includegraphics[width=7cm,angle=270]{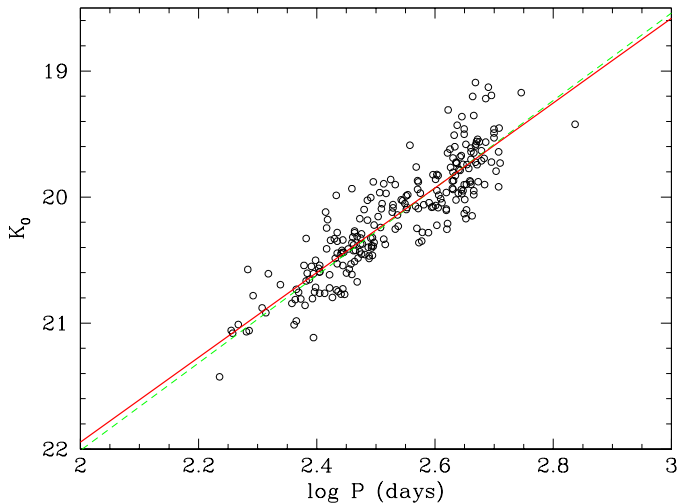}
  \caption[]
	{Mira PL relation in NGC~5128 for variables with best
	determined periods and colors bluer than $J-K \leq 1.4$. 
	The solid red line is our best fit to the Mira
	sequence. The dotted green line is  the 
	Feast et al.~(\cite{feast+02})
	relation.}
  \label{plbest}
\end{figure}  

A well-defined PL relation has been found for Miras in
the Large Magellanic Cloud (Glass \& Lloyd Evans~\cite{glass&lloydevans81},
Wood~\cite{wood+99}), the Small Magellanic Cloud (Cioni et
al.~\cite{cioni+03}), the Galactic Bulge (Glass et al.~\cite{glass+95}),
the solar neighbourhood (van Leeuwen et al.~\cite{vanleeuwen+97})
and in Galactic globular clusters (Feast et al.~\cite{feast+02}).  The
relation holds for both ${\rm M}_{\rm bol}$ and ${\rm M}_K$.  Since Miras are
very luminous, their tight PL relation makes them interesting for distance
determination to other galaxies.

Calibration of the $K$-band PL relation relies on the LMC PL 
relation for Miras.
The Feast et al.~(\cite{feast+89}) fit to the LMC Mira PL relation is:
\begin{equation}
M_K = -3.47 \log P + \beta
\end{equation}
The zero point of $\beta=0.91$ has been derived by Feast
(\cite{feast96}) using LMC Miras. More recently  
Whitelock et al.~(\cite{whitelock+00}) 
used Hipparcos parallaxes for Solar neighbourhood
Miras to derive $\beta=0.84 \pm 0.14$ and Feast et al.~(\cite{feast+02}) 
obtained $\beta=0.93 \pm 0.14$ for
globular cluster Miras using the cluster distances determined from
subdwarf fitting with Hipparcos parallaxes of subdwarfs
(Caretta et al.~\cite{caretta+00}). The straight mean
of the two Hipparcos based zero points
yields $\beta=0.88 \pm 0.10$ and a Large Magellanic Cloud distance modulus
of $18.60 \pm 0.10$. Preferring instead a somewhat shorter LMC distance, 
$(m-M)_0^{LMC}=18.50 \pm 0.04$ (Alves et al.~\cite{alves+02}), I use 
a zero
point of $\beta=0.98 \pm 0.11$ (where uncertainty in the LMC distance
modulus has been added to the
zero point error in quadrature).

Glass \& Lloyd Evans (\cite{glass+lloydevans03}) have
re-reanalysed the Mira PL relation using the MACHO data for the same stars
as Feast et al.~(\cite{feast+89}) in the LMC. They conclude that the periods
of these variables did not change significantly and remained essentially
constant over 2--3 decades. Their best fit to the $K$-band Mira PL relation
is:
\begin{equation}
K = -3.52(\pm 0.21) \log P + 19.64 (\pm 0.49) ~~ \sigma=0.13
\end{equation}

In order to determine the distance to NGC~5128 only Miras with colors 
$J_s-K_s<1.4$ have been selected. This ensures that reddening from 
circumstellar shells that may be present
in redder LPVs would not dim the $K$-band magnitudes and also matches the 
colors of Mira variables used to derive the Mira PL relation in the LMC
($(J-K)<1.5$, Feast et al.~\cite{feast+89}). 
Moreover, only those LPVs that had the
most regular periods, and could be fitted with a sine-curve with 
$\chi^2<5$ were chosen. They are plotted in Fig.~\ref{plbest} 
together with the best linear least-square 
fit to the data (thick solid line):
\begin{equation}
K_0=-3.37(\pm 0.11) \log P + 28.67(\pm 0.29)~~ (N=240)~~ 
\sigma=0.20
\end{equation}
The fits to Field~1 and Field~2 LPVs are identical within the errors. 
Largely, the error in the constant term is due to the extrapolation 
of the relation to the period of 1~day. The mean $\log P$ is $\sim 2.5$. 
If instead the fit is made at the mean period with the following 
functional form:
\begin{equation}
K_0 = A (\log P - X) + B
\end{equation}
the error of the constant term drops to only $0.02$. 

The slope of the PL relation in 
NGC~5128 agrees well with that in the LMC.
It is within $1\sigma$ of the slope of  the 
Feast et al.~(\cite{feast+89}) relation and $1.5 \sigma$ of 
the Glass \& Lloyd Evans (\cite{glass+lloydevans03}) PL relation. 
This is yet another piece of evidence that the slope of the PL relation
in the $K$-band is universal,
lending confidence to the distance determination through the PL relation.

In order to determine the distance modulus of NGC~5128 from the Mira
PL relation, one has to assume that the slope is universal and use the 
same value as in the calibrating PL relation (i.e. that in the LMC). 
It is straightforward then to calculate the distance modulus from the 
zero point difference of the two PL relations.
Restricting the slope to $-3.47$, and determining the zero-point of 
the relation at $(\log P-2.3)$, approximately the mean period of the 
LMC stars used to determine the relation (Feast et al.~\cite{feast+89}, 
Glass \& Lloyd Evans \cite{glass+lloydevans03})
the best fitting zero point is $28.94 \pm 0.03$ (dotted green line
in Fig.~\ref{plbest}) with the same RMS of the fit. Using the above 
discussed zero point of 
$\beta= 0.98$, I derive  a distance modulus to NGC~5128 of 
$(m-M)_0=27.96 \pm 0.11$.
If instead the slope of the PL relation is fixed to $-3.52$ 
(Glass \& Lloyd Evans \cite{glass+lloydevans03}), the resulting distance 
modulus to NGC~5128 is $(m-M)_0=27.93 \pm 0.11$.

Feast et al.~(\cite{feast+89}) as well as Glass \& 
Lloyd Evans (\cite{glass+lloydevans03})
noted the group of stars with periods in excess of $\sim 420$ 
days which lie some $\sim 0.7$ mag
above the PL relation. Due to the fact that  a few of them have been 
found to be Li-rich, it was suggested that they are more massive, hence 
younger and are in the Hot-Bottom Burning 
phase  (Smith et al.~\cite{smith+95}). 
So, restricting the LPV data in NGC~5128 to LPVs with periods 
shorter than 400 days, where the above mentioned LMC PL relations have 
been calibrated, distance moduli of
$27.95 \pm 0.12$ and $27.92 \pm 0.12$ are derived adopting slopes of 
$-3.47$ and $-3.52$, respectively. These distance moduli are 
virtually indistinguishable from that determined 
from the full data set and they point out that most of the long period 
stars do follow the same Mira PL relation as shorter period Miras. 
They are not overluminous and in the hot-bottom burning phase. 

There are a few dozen stars in Fig.~\ref{plall} that are
$\ga 2 \sigma$ above 
the PL relation. Some of them might be brighter due to contributing 
light from their neighbours, but some show indications of humps in 
their light-curves. Similar humps have been found in MACHO 
light curves of the LMC Li-rich stars by Glass \& Lloyd Evans 
(\cite{glass+lloydevans03}). However, before being able to place firm 
conclusions, accurate light curves over several periods should be 
determined for a larger statistical sample of Li-rich hot-bottom burning
AGB stars.

%
%
\section{Conclusions}
\label{conclusions}

The tip of the RGB was detected in Field~1 at $K_s=21.24 \pm 0.05$, yielding 
a distance
modulus of NGC~5128 of $(\mathrm{m}-\mathrm{M})_0=27.87 \pm 0.16$. 
The comparison of the $H$-band RGB tip luminosity in the Galactic Bulge
and NGC~5128 implies a similar distance modulus (Table~\ref{tabdist2}).
The RGB tip in Field~2 is not a sharp feature due to a
brighter completeness limit, 
larger photometric errors and the presence of blends and AGB stars. 

A large population of stars above the tip of the RGB contains 2176 stars
in the outer halo field (Field~1) and 6072 stars in the inner halo
field (Field~2). Subtracting  these 
foreground sources, detected LPVs, as
well as maximum probable number of blends of two RGB tip stars, 
there are some 1150 and 150 
non-variable stars brighter than the first ascent giant branch tip
in the two fields. LPVs account for 26\% and 70\%
of the AGB population in Fields~1 and 2, respectively. The high
luminosity ($M_K\leq-8.7$) achieved by AGB stars 
is a sign of an intermediate-age population. 

\begin{table}
\centering
\caption[]{Summary of the distance determinations in this work.}
\label{tabdist2}
\begin{tabular}{cll}
\hline
\hline
DM (mag) & Method & Field\\
\hline
\hline
$27.87 \pm 0.16$& $K$-band RGB tip &F1\\
$27.9 \pm 0.2$& $H$-band RGB tip &F1 \\
$27.96 \pm 0.11$& $K$-band Mira PL relation &F1 \& F2 \\
\hline
\end{tabular}
\end{table}

The first Mira period-luminosity relation outside the Local Group is
presented. Miras with the best determined and most regular periods that do not
have red colors ($J_s-K_s<1.4$) were used to 
determine the distance of NGC~5128 from a period-luminosity relation. 
I derive the distance modulus
of $27.96 \pm 0.11$, adopting the LMC distance modulus of 18.50 
(Alves et al.~\cite{alves+02}). 

The mean distance of the Miras, obtained from fields on both sides of the
center of NGC~5128, is slightly larger than the mean distance of the
red giants in Field~1. This indicates that the orientation of NGC~5128 is 
such that the north-eastern part of the galaxy is closer to us. 
However, the relative distance of the 
two fields is smaller than $\sim 0.03$~Mpc and negligible compared to 
the distance from the Milky Way.
The mean of the two methods yields a distance to NGC~5128 of 
$(m-M)_0=27.92 \pm 0.19$ ($D=3.84 \pm 0.35$ Mpc), very close to the
$(m-M)_0=27.98\pm 0.14$ measurement of 
Harris et al.~(\cite{harris+99}) and within $1\sigma$ of the distance
modulus $27.8 \pm 0.2$ result of Soria et al.~(\cite{soria+96}). 

%
%

\begin{acknowledgements}

I would like to thank M.-R. Cioni for useful discussions, 
M. Feast for suggestions regarding 
the Mira PL relation and to Chris Lidman for communicating the 
transformation equations between different near-IR photometric systems.
Thanks also go to Tim Bedding for carefully reading the manuscript
and for making a number of useful comments. Comments from 
H. Ford and the second anonymous referee improved the clarity of 
the presentation.
I am indebted to many ESO staff 
astronomers who took the data presented in this
paper in service mode operations at Paranal Observatory.
\end{acknowledgements}

\end{document}